\documentclass[pdfa,a4paper,USenglish,cleveref,autoref,thm-restate]{lipics-v2021}

\pdfoutput=1 

\hideLIPIcs

\nolinenumbers 
\usepackage{tikz}
\usetikzlibrary{positioning}
\usetikzlibrary{calc}

\title{A Formal Proof of Complexity Bounds on Diophantine Equations}

\author{Jonas Bayer}{Department of Pure Mathematics and Mathematical Statistics, University of Cambridge, UK}{jcb234@cam.ac.uk}{https://orcid.org/0009-0006-2500-3146}{}
\author{Marco David}{Department of Physics, University of California, Berkeley, CA, USA}{mdv@berkeley.edu}{https://orcid.org/0000-0001-5673-5389}{}

\authorrunning{J. Bayer and M. David}
\Copyright{Jonas Bayer and Marco David}



\supplementdetails[cite={dagstuhl-artifact-24709},subcategory={Formal Proof Development}, linktext={https://www.isa-afp.org/ \\entries/Diophantine\_Universal\_Pairs.html}]{Software}{https://www.isa-afp.org/entries/Diophantine_Universal_Pairs.html}

\acknowledgements{This project would not exist without Yuri Matiyasevich who proposed the formalization of Hilbert's Tenth Problem and encouraged our exploration of Universal Pairs. We are deeply grateful to Dierk Schleicher for his unwavering support throughout, both on a personal level and through his mathematical insight. We thank Malte Haßler, Thomas Serafini and Simon Dubischar for their mathematical contributions to the project. We are indebted to the group of students who contributed at various stages of the formalization. In particular, we thank Timothé Ringeard and Xavier Pigé for leading the formal proof of the Bridge Theorem; Anna Danilkin and Annie Yao for their work on \texttt{poly\_extract}, \texttt{poly\_degree} and further developments in multivariate polynomial theory; Mathis Bouverot-Dupuis, Paul Wang, Quentin Vermande, and Theo André for their formalization of coding theory; Loïc Chevalier, Charlotte Dorneich, and Eva Brenner for the formalization of Matiyasevich’s polynomial and relation combining; and Zhengkun Ye and Kevin Lee for their contributions to formal degree calculations. The above highlights their primary contributions, though many have supported other aspects of this work whenever they were able. We also thank all other members of the student work group for their involvement in the project. We gratefully acknowledge support from the D\'epartement de math\'emathiques et applications at \'Ecole Normale Sup\'erieure de Paris.}

\relatedversiondetails{Full Version}{https://arxiv.org/abs/2506.20909}

\usepackage{needspace}
\newcommand{\N}{{\mathbb N}}
\newcommand{\Z}{{\mathbb Z}}
\newcommand{\B}{\mathscr{B}}
\newcommand{\thmIdec}{\mathscr}
\newcommand{\X}{\thmIdec X}
\newcommand{\K}{\thmIdec K}
\newcommand{\Y}{\thmIdec Y}
\newcommand{\bLowercase}{\mathfrak{b}}

\DeclareMathOperator{\coeffs}{coeffs}

\definecolor{isarblue}{HTML}{006699}
\definecolor{isargreen}{HTML}{009966}
\lstdefinelanguage{isabelle}{keywords=[1]{type_synonym,datatype,fun,abbreviation,definition,proof,lemma,theorem,corollary,by,have,from,with,using,lift_definition,locale,consts},
keywordstyle=[1]\bfseries\color{isarblue},
keywords=[2]{fixes,defines,where,assumes,shows,and,is},
keywordstyle=[2]\bfseries\color{isargreen},
keywords=[3]{if,then,else,case,of,SOME,let,in,O},
keywordstyle=[3]\color{isarblue},
commentstyle=\color{gray},
morecomment=[l]{(*}
}
\newcommand{\formalref}[2]{\href{https://gitlab.com/hilbert-10/universal-pairs/-/blob/cd0c7d8a9d1a08f2a3a565b009929ab6b43bfbd7/#1\#L#2}{#1 (line #2)}}

\def\topmattervskip{0.34}

\EventEditors{Yannick Forster and Chantal Keller}
\EventNoEds{2}
\EventLongTitle{16th International Conference on Interactive Theorem Proving (ITP 2025)}
\EventShortTitle{ITP 2025}
\EventAcronym{ITP}
\EventYear{2025}
\EventDate{September 28 -- October 1, 2025}
\EventLocation{Reykjavik, Iceland}
\EventLogo{}
\SeriesVolume{352}
\ArticleNo{3}

\begin{document}

\maketitle

\begin{abstract}
We present a universal construction of Diophantine equations with bounded complexity in Isabelle/HOL. This is a formalization of our own work in number theory~\cite{manuscript}.

Hilbert's Tenth Problem was answered negatively by Yuri Matiyasevich, who showed that there is no general algorithm to decide whether an arbitrary Diophantine equation has a solution. However, the problem remains open when generalized to the field of rational numbers, or contrarily, when restricted to Diophantine equations with bounded complexity, characterized by the number of variables $\nu$ and the degree $\delta$. If every Diophantine set can be represented within the bounds $(\nu, \delta)$, we say that this pair is \emph{universal}, and it follows that the corresponding class of equations is undecidable. In a separate mathematics article, we have determined the first non-trivial universal pair for the case of integer unknowns. 

In this paper, we contribute a formal verification of this new result. In doing so, we markedly extend the Isabelle AFP entry on multivariate polynomials~\cite{polynomials-afp}, formalize parts of a number theory textbook~\cite{nathanson}, and develop classical theory on Diophantine equations~\cite{MR75} in Isabelle. In addition, our work includes metaprogramming infrastructure designed to efficiently handle complex definitions of multivariate polynomials. Our mathematical draft has been formalized while the mathematical research was ongoing, and benefited largely from the help of the theorem prover. We reflect on how the close collaboration between mathematician and computer is an uncommon but promising modus operandi.
\end{abstract}

\section{Introduction}
\subsection{Hilbert's Tenth Problem}
The tenth problem on David Hilbert's 1900 list of twenty-three mathematical problems asks for an algorithm to decide the solvability of any arbitrary Diophantine equation~\cite{hilbert-problems}. Like many problems on his list, Hilbert's Tenth Problem has provided inspiration for much mathematical research in the last century, and still does so until today. While the original problem was solved more than fifty years ago~\cite{exp-diophantine}, this paper presents the formal verification of results related to a new quantitative theorem, stronger than the original solution.

A \emph{Diophantine equation} is a multivariate polynomial equation over the integers. There are countably infinitely many such equations, and they form a very general language: many difficult problems across number theory and related fields can be equivalently expressed as a single Diophantine equation~\cite{riemann-register-machine}. One often considers parametric equations of the type $P(a, z_1, \ldots, z_\nu)=0$ where $a$ is called the parameter and $z_1, \ldots, z_\nu$ are called unknowns. Then a \emph{Diophantine set} $A \subset \N$ is defined as any set of parameters such that $a \in A$ if and only if there are unknowns $z_1, \ldots, z_\nu \in \Z$ with $P(a, z_1, \ldots, z_\nu) = 0$.

The Davis-Putnam-Robinson-Matiyasevich (DPRM) theorem (1970)~\cite[Chapter~1]{matiyasevich-book} states that every recursively enumerable set is Diophantine; this is a deep relation between number theory and computability theory. An enumeration argument shows that, conversely, every Diophantine set is recursively enumerable. The resulting equivalence implies that there are Diophantine sets which are not decidable, and hence proves that Hilbert's Tenth Problem is undecidable.

\subsection{Universal Diophantine Equations and Pairs}
Because the set of all recursively enumerable (r.e.) sets is itself recursively enumerable, there must exist a corresponding Diophantine equation for this set, representing all r.e.\ sets, i.e.\ all Diophantine sets at once. An equation with this property is known as a \emph{universal Diophantine equation}, and an explicit example can be found in an article by Jones~\cite{9var}.

It is now natural to ask about the minimal complexity required for such universal equations. A popular measure of complexity is the number of unknowns $\nu$ required and the total degree~$\delta$ of the polynomial. If all Diophantine sets can be represented by a polynomial in at most $\nu$ unknowns and in degree at most $\delta$, we call $(\nu, \delta)$ a \emph{universal pair}. There is a trade-off between the two measures: one may reduce the degree by introducing auxiliary unknowns, and vice versa eliminate unknowns at the cost of increasing the degree.

There is a subtle but interesting distinction between allowing the unknowns $z_1, \ldots, z_\nu$ of $P$ to take values in the natural numbers, including zero, versus in the integers. Universal pairs can be defined for both cases, and we shall write $(\nu, \delta)_\N$ and $(\nu, \delta)_\Z$ to make the distinction explicit. Previous work has primarily focused on universal pairs over the natural numbers.
In this case, it was Matiyasevich~\cite{M77} who first announced the smallest currently known number of unknowns $\nu = 9$. Jones~\cite{9var} later published a complete proof that $(9, 1.64 \cdot 10^{45})_\N$ is universal, together with the lowest currently known degree $\delta = 4$. More precisely, the pair $(58, 4)_\N$ is universal, too.

\enlargethispage{\baselineskip}
Our work contributes the first derivation of a non-trivial universal pair over the integers. Note that any universal pair over $\N$ induces a pair over $\Z$. Every $n \in \N$ can be written as $x^2 + y^2 + z^2 + w^2$ for integers $x,y,z,w \in \Z$ using Lagrange's four square theorem. Thus, a Diophantine equation with unknowns in $\N$ can be transformed into an equation with unknowns in $\Z$ by introducing additional unknowns. However, this conversion gives bounds that are far from optimal. 

Using the four-squares substitution for $\nu = 9$ over the natural numbers would thus require 36 integer unknowns. A more efficient substitution is possible using the more sophisticated three squares theorem (cf. \cref{sec:three-squares}); this would yield universality for 27 integer unknowns. 

\subsection{First Nontrivial Bounds over \texorpdfstring{\boldmath $\Z$}{the Integers}}

In this article, we formalize the following improvement to only eleven unknowns:

\begin{theorem}\label{thm:11-universal-pair}
Let $(\nu, \delta)_\N$ be universal. Then
\[ \big(11, \eta(\nu, \delta) \big)_\Z \]
is universal where $\eta(\nu, \delta) = 15 \, 616 + 233\,856 \; \delta + 233\,952 \; \delta \, (2 \delta + 1)^{\nu+1} + 467\,712 \; \delta^2 \, (2 \delta + 1)^{\nu+1}$.
\end{theorem}
In particular, 
\[
(11, 1.68105\cdot 10^{63}) \qquad \text{and} \qquad (11, 9.50818 \cdot 10^{53}) 
\]
are universal over $\Z$.

This theorem has been formalized in Isabelle/HOL and is the focus of this paper. The technique used to reduce the number of unknowns to eleven was developed by Sun~\cite{Sun}. However, that work does not consider the degree of the constructed polynomials.

Observe that, in principle, the second pair stated above is better than the first. We state both because the second pair depends on Jones' universal pair $(32, 12)_\N$ of which there is no published proof in the literature. The first pair depends instead on $(58, 4)_\N$.

\cref{thm:11-universal-pair} implies the following stronger response to Hilbert's Tenth Problem.
\begin{corollary}\label{cor:strong-htp}
Consider the class of Diophantine equations with at most 11 unknowns and at most degree $1.69 \cdot 10^{63}$. Hilbert's Tenth Problem is unsolvable for this class: There is no algorithm that can determine for all equations in this class whether they have a solution or~not. 
\end{corollary}

Theorem~\ref{thm:11-universal-pair} crucially rests on the following result.

\begin{theorem}\label{thm:11-unknowns}  
Let $P \in \Z[a, z_1, \ldots, z_\nu]$. Then there is an explicit construction of a polynomial $\tilde Q \in \Z[a, z_1, \ldots, z_{11}]$ such that, for all $a \in \N$,
\[
\exists z_1, \ldots, z_\nu \in \Z^\nu. \; P(a, z_1, \ldots, z_\nu) = 0 \iff \exists \tilde z_1, \ldots, \tilde z_{11} \in \Z^{11}. \; \tilde Q(a, \tilde z_1, \ldots, \tilde z_{11}) = 0.
\]
\end{theorem}

These theorems and the corollary appear together in our recent manuscript~\cite{manuscript} which has been written in parallel to the development of the formal proof presented here.

\section{Related Work}
Hilbert's Tenth Problem and related statements have been formalized in a variety of different systems. Note that there are many different but equivalent options to define recursively enumerable sets, either algebraically or by using some Turing-complete model of computation. In the following, we summarize formalizations of results related to Hilbert's Tenth Problem, highlighting the different choices that the authors have made.

In Lean, Carneiro~\cite{lean-formalization} has formalized that exponentiation is Diophantine, which has historically been the last step to be completed in the proof of the DPRM theorem. In Coq, Larchey-Wendling and Forster~\cite{coq-formalization} have formalized the complete DPRM theorem, elegantly using Conway's FRACTRAN language as a Turing-complete model of computation. They moreover formalize a reduction of their library of unsolvable problems, including two variants of the Halting Problem, to Hilbert's Tenth Problem. In Mizar, Pak~\cite{mizar-formalization} has formalized the DPRM theorem, defining recursively enumerable sets by Davis' normal form. In Isabelle, we have formalized the DPRM theorem using register machines as a model of computation~\cite{isabelle-formalization}.

\medskip

Mathematically, we rely on techniques originally developed for working with Diophantine equations with natural number unknowns; these have been developed by Matiyasevich, Robinson~\cite{MR75} and Jones~\cite{9var}. Over the integers, Sun~\cite{Sun}
first announced many results. In particular, he proved that any Diophantine equation can be reduced to an equivalent one with only eleven integer unknowns. 

Unfortunately, the published literature often lacked sufficient detail to formalize, and sometimes even to understand, the existing proofs. As such, our detailed mathematical manuscript~\cite{manuscript} now reproduces many of the results and proofs cited above. The improved clarity in exposition is a notable product of this formalization. 

\section{Preliminaries}

\subsection{Multivariate Polynomials in Isabelle}
Throughout our work, we rely on the formalization of executable multivariate polynomials by Sternagel et al.~\cite{polynomials-afp}. The type of multivariate polynomials is defined as
\begin{align}\label{eq:typedef-mpoly}
\alpha\; \texttt{mpoly} := (\N \Rightarrow_0 \N) \Rightarrow_0 \alpha
\end{align}
for any type $\alpha$ with a zero. Most commonly we will take $\alpha = \Z$ or a unital, commutative ring. We use $\beta \Rightarrow_0 \alpha$ to denote a \emph{finite assignment}, i.e.\ a function from $\beta$ to $\alpha$ that takes only finitely many nonzero values. Each finite assignment $m : \N \Rightarrow_0 \N$ defines an abstract multivariate monomial $\prod_i x_i^{m(i)}$, giving the exponents of each $x_i$, for $i\in \N$. A multivariate polynomial then assigns a coefficient from $\alpha$ to each monomial. Of course, this type inherits much algebraic structure from the underlying coefficient type: if $\alpha$ is a ring or an integral domain, then so is $\alpha$ \texttt{mpoly}. With this formalization, we extend the existing library by well over 2,000 lines of proof.

A key first addition to the existing library is an explicit expansion, for arbitrary $\nu \in \N$, of a general multivariate polynomial $P \in \alpha[z_0, \ldots, z_\nu]$ as
\begin{equation}\label{eq:multivariate-expansion}
P(z_0, \ldots, z_\nu) = \sum_{i_0, \ldots, i_\nu \, \in \, \N} c_{i_0, \ldots, i_\nu} \prod_{s=0}^\nu z_s^{i_s}.
\end{equation}
This expression is a polynomial as long as only finitely many coefficients $c_{i_0, \ldots, i_\nu}$ are non-zero. Below, we abbreviate the indices with a single multi-index $\mathbf{i} = (i_0, \ldots, i_\nu) \in \N^{\nu+1}$, defining $\|\mathbf{i}\| := \sum_{s=0}^\nu i_s$.

With $\delta := \deg P$ and the constraint $\|\mathbf{i}\| \leq \delta$, we can reformulate the same expression as a finite sum, which is much better behaved for formal proofs in Isabelle.
\begin{equation}\label{eq:multivariate-expansion-bounded}
P(z_0, \ldots, z_\nu) = \sum_{\substack{\mathbf{i} \in \N^{\nu+1} \\ \|\mathbf{i}\| \le \delta}} c_{i_0, \ldots, i_\nu} \prod_{s=0}^\nu z_s^{i_s}.
\end{equation}

Even if we will not provide proofs of equations~\eqref{eq:multivariate-expansion} and~\eqref{eq:multivariate-expansion-bounded}, neither equation follows trivially from the definition~\eqref{eq:typedef-mpoly} of the type $\alpha$ \texttt{mpoly}. In Isabelle, we have formalized equation~\eqref{eq:multivariate-expansion-bounded} as follows. The term \texttt{Const C} represents a constant polynomial with value $C$, and the term \texttt{Var s} represents the linear polynomial $z_s$. The term \texttt{max\_vars P} gives the highest variable index that appears in $P$, denoted $\nu$ above.

\begin{lstlisting}[caption=\formalref{MPoly\_Utils/Poly\_Expansions.thy}{527}.]
lemma mpoly_multivariate_expansion:
  fixes P :: "$\alpha$ mpoly"
  shows "P = $\sum$i | length i = max_vars P + 1 $\wedge$ ||i|| $\leq$ total_degree P.
          Const (coeff P i) * ($\prod$s = 0..max_vars P. (Var s) ^ (i ! s)))"
\end{lstlisting}

\begin{remark}
In the interest of legibility, throughout the paper, we are presenting Isabelle-style pseudo-code in which we have dropped some notation that is syntactically but not semantically necessary.
Moreover, by writing \texttt{||i||} above we mean the library function \texttt{sum\_list i}. We write $\alpha$ for the free type variable \texttt{'a} representing the ring of coefficients. We assume $\alpha$~\texttt{::} \texttt{comm\_ring\_1}, a commutative and unital ring, throughout.
\end{remark}

In the same way, we developed other basic notation and tools to efficiently reason about the type $\alpha$ \texttt{mpoly}. For example, we formalize that if one type injects into a larger type like $\Z \hookrightarrow \mathbb{Q}$, then the same holds for $\Z$ \texttt{mpoly} $\hookrightarrow \mathbb{Q}$ \texttt{mpoly}. We also prove their correct behavior with respect to standard operations on polynomials: addition, multiplication, exponentiation et cetera.

\paragraph*{Substitution}
One property that has been vital for the iterative definition of universal Diophantine constructions is the substitution of polynomials into polynomials. The existing AFP entry provides a morphism \texttt{reduce\_nested\_mpoly ::} \texttt($\alpha$ \texttt{mpoly) mpoly} $\Rightarrow$ $\alpha$ \texttt{mpoly}. However, we instead chose to formalize substitutions more directly and more explicitly, adapted to the constructive proof presented in the next section.

\begin{lstlisting}[caption=\formalref{MPoly\_Utils/Substitutions.thy}{11}.]
definition poly_subst_monom :: "(nat$\;\Rightarrow\;$$\alpha\;$mpoly) $\Rightarrow$ (nat$\;\Rightarrow_0\;$nat) $\Rightarrow$ $\alpha\;$mpoly"
  where "poly_subst_monom f m = $\prod$s. (f s) ^ (m s)"

definition poly_subst :: "(nat$\;\Rightarrow\;\alpha\;$mpoly) $\Rightarrow$ $\alpha\;$mpoly $\Rightarrow$ $\alpha\;$mpoly"
  where "poly_subst f P = $\sum$m. (Const (coeff P m)) * poly_subst_monom f m"
\end{lstlisting}
This formalizes an expansion of the polynomial $P$ in analogy to the above lemma, except that each \texttt{Var s} has been replaced by corresponding substitution $f_s$, which is now allowed to be any arbitrary multivariate polynomial. 

In standard mathematical notation, given a polynomial $P \in \Z[z_0, \ldots, z_\nu]$, we would write this substitution as a list of $(\nu+1)$-many polynomials $f = (f_0, \ldots, f_\nu) \in \Z[z_0, \ldots, z_\nu]^{\nu+1}$. One often requires the corresponding correctness statement: the order of evaluation does not matter. That means, given a vector of integers $\mathbf{z} = (z_0, \ldots, z_\nu) \in \Z^{\nu+1}$, we need to prove that
\begin{equation}\label{eq:substitution-evaluation}
P(f)(\mathbf{z}) = P(f(\mathbf{z})),
\end{equation}
where $f(\mathbf{z}) = (f_0(\mathbf{z}), \ldots, f_\nu(\mathbf{z})) \in \Z^{\nu+1}$ is the evaluation of each substitution polynomial at~$\mathbf{z}$. In Isabelle, the evaluation map is formalized as \texttt{insertion z P} for an assignment of the type \texttt{z :\!\!\! nat $\Rightarrow \alpha$} and a multivariate polynomial \texttt{P :\! $\alpha$ mpoly}. Then, equation~\eqref{eq:substitution-evaluation} reads:
\begin{lstlisting}[caption=\formalref{MPoly\_Utils/Substitutions.thy}{82}.]
lemma insertion_poly_subst:
  "insertion z (poly_subst f P) = insertion ((insertion z) $\circ$ f) P"
\end{lstlisting}

\paragraph*{Number of Variables and Total Degree}
To explicitly formalize universal pairs, we need to be able to calculate the degree and number of variables of a multivariate polynomial in Isabelle. The AFP entry~\cite{polynomials-afp} defines both \texttt{vars} and \texttt{total\_degree}, to which we have added the definition \texttt{max\_vars}.
\begin{lstlisting}[caption={Excerpt from the AFP entry on multivariate polynomials~\cite{polynomials-afp}}.]
lift_definition vars :: "$\alpha$ mpoly $\Rightarrow$ nat set"
  is "$\lambda$P. $\bigcup$ (keys ` keys P)"

lift_definition total_degree :: "$\alpha$ mpoly $\Rightarrow$ nat"
  is "$\lambda$P. Max (insert 0 (($\lambda$m. sum (lookup m) (keys m)) ` keys P))"
\end{lstlisting}
\begin{lstlisting}[caption=\formalref{MPoly\_Utils/Variables.thy}{185}.]
definition max_vars :: "$\alpha$ mpoly $\Rightarrow$ nat"
  where "max_vars P $\equiv$ Max (insert 0 (vars P))"    
\end{lstlisting}
The function \texttt{keys} is defined to be the set of all elements of the domain which are mapped to a non-zero value. In particular, \texttt{keys P} is the set of all monomials which appear in $P$ with non-zero coefficient. The map \texttt{lookup} denotes function evaluation for finite assignments. Recall also that \texttt{sum f S} $= \sum_{s\in S} f(s)$.

In Isabelle, the \texttt{Max} of a set is not defined for empty sets. Because any constant polynomial has \texttt{vars (Const C)} $= \emptyset$, inserting the number $0$ in the definition of \texttt{max\_vars} ensures that the set is always non-empty. When constructing polynomials we always make sure to not skip any variable indices. Moreover, because we work with parametric polynomials, the zeroth variable is always reserved for the parameter $a$. Accordingly, the above definition has the correct semantics for this project.

The definition of \texttt{total\_degree} works in analogy to \texttt{max\_vars}. Because the zero polynomial has \texttt{keys (Const 0)} $= \emptyset$, we again include \texttt{insert 0}. However, this formalizes the degree of the zero polynomial as~$0$. A consequence of this definition is, for example, that the assumptions $P \neq 0$ and $Q \neq 0$ are now required for the lemma \texttt{total\_degree (P * Q) = total\_degree P + total\_degree Q}.

\medskip

We formalize appropriate simplification and correctness properties of these definitions with respect to all other functions that we use to construct polynomials. This includes exponentiation, finite sums and products, type casting, and substitutions. Because we are concerned with upper bounds of the degree only, it suffices to establish inequalities, which require fewer assumptions. Indeed, \texttt{total\_degree (P * Q) $\leq$ total\_degree P + total\_degree Q} is true without any assumptions on $P$ or $Q$.

\begin{remark}
With the many substitutions of polynomials into polynomials required for our construction, it is useful to clearly state how the calculations of \lstinline{max_vars} and \lstinline{total_degree} interact with iterated substitutions. For the variables, it is not hard to prove the following, which can be applied recursively.
\begin{lstlisting}[caption=\formalref{MPoly\_Utils/Substitutions.thy}{316}.]
lemma vars_poly_subst:
  "vars (poly_subst f P) $\subseteq$ $\bigcup$s $\in$ vars P. vars (f s)"
    \end{lstlisting}
\enlargethispage{-.3\baselineskip}
However, for the degree, it becomes necessary to fully expand all substitutions to obtain tight bounds. For example, consider first evaluating the degree of $P = Q_1 Q_2 + Q_3$. Given an arbitrary substitution $f$, this polynomial can have many different degrees in the variables~$z_s$.\footnote{Consider, for example, the two substitutions $f = (Q_1, Q_2, Q_3) = (z_1, z_2, z_2^2)$ and $f = (z_1^5, z_2^3, z_2)$.} Of course, there is an explicit formula for the total degree of this $P$. However, for both arbitrary $f$ and arbitrary $P$, an elementary recursive step, analogous to the above lemma for variables, cannot suffice.

Instead, we rely on the following auxiliary definition, which will be used by the automation tools introduced below. We abbreviate \lstinline{Max (insert 0 -)} by just \lstinline{Max0 -}.
\begin{lstlisting}[caption=\formalref{MPoly\_Utils/Total\_Degree\_Env.thy}{12}.]
lift_definition total_degree_env :: "(nat $\Rightarrow$ nat) $\Rightarrow$ $\alpha$ mpoly $\Rightarrow$ nat"
  is "$\lambda$e P. Max0 (($\lambda$m. sum ($\lambda$i. e i * lookup m i) (keys m)) ` (keys P)))"
    \end{lstlisting}
Here, $e : \N \rightarrow \N$ denotes an ``environment'' of degrees $\delta_s$ for each basic variable $z_s$. Note the special case \lstinline{total_degree_env ($\lambda$_. 1) P = total_degree P}. The environment is updated with the degree of each $f_s$ whenever we encounter \lstinline{poly_subst f P}, modeling our algorithm to compute the degree of an arbitrary multivariate polynomial. 
\end{remark}

\paragraph*{Custom Metaprogramming Infrastructure}
Recall that \cref{thm:11-universal-pair,thm:11-unknowns} assert that any Diophantine equation can be expressed as a polynomial $\tilde Q$ with only eleven integer unknowns. Consequently, the natural choice for the type of $\tilde Q$ in Isabelle is \texttt{int mpoly}. Since $\tilde Q$ is constructed by substituting various previously defined polynomials into each other, one might expect that these intermediate polynomials also have the type \texttt{int mpoly}.

However, this approach presents notable difficulties when formalizing the \cref{thm:coding,thm:lucas} which will be presented in the following section. Instead of using meaningful variable names like $a, f, g$, one must work with \texttt{Var 0}, \texttt{Var 1}, \texttt{Var 2} and frequently refer to the insertion map explicitly. More critically, substituting polynomials into one another requires care to avoid collisions of variable indices. These challenges cannot be fully mitigated through new notation alone.

To address these issues, this project developed a custom metaprogramming framework that effectively bridges Isabelle terms and the type \texttt{int mpoly}. This allows our formalization to benefit from the best of both worlds, significantly simplifying our proofs. 

Using this framework, a polynomial $P$ is initially formalized as a standard Isabelle term of type \lstinline{P :: int => int => ... => int}. The command \texttt{poly\_extract P}, implemented by Anna Danilkin in our workgroup, then generates a new term \texttt{P\_poly} of type \texttt{int mpoly} and automatically proves a lemma \texttt{P\_correct}. This lemma guarantees that evaluating \texttt{P\_poly} yields the same result as evaluating the original function \texttt{P}. The extraction succeeds if \texttt{P} is a valid polynomial: constructed from integer constants and variables, addition, multiplication, exponentiation with a natural number, and any term for which appropriate translations to a polynomial already exist from previous applications of \texttt{poly\_extract}. 

\medskip

Calculating the total degree of the final universal Diophantine equation $\tilde Q$ and proving its correctness also posed significant challenges due to the depth of iterated substitutions. The correctness proofs, in particular, require thousands of invocations of the various simplification lemmas $\deg(PQ) \leq \deg P + \deg Q$ for multiplication, $\deg(P + Q) \leq \max\{\deg P, \deg Q\}$ for addition, and $\deg P^n \leq n \deg P$  --   specifically, one invocation for each arithmetic operation that appears in our construction. To automate both the calculation of the degree and an accompanying proof of correctness, Annie Yao developed the command \lstinline{poly_degree}.

When invoked on a term of type \lstinline{P_poly :: int mpoly}, this command recursively computes (an upper bound for) the degree of this term, called \lstinline{P_poly_degree}, and automatically proves a lemma \texttt{P\_poly\_degree\_correct}. The lemma states that the total degree of \texttt{P\_poly} is indeed bounded above by \lstinline{P_poly_degree}. The calculation and proof succeed when \texttt{P\_poly} is constructed from basic arithmetic operators, variables,  constants (including arbitrary \mbox{Isabelle} terms of type \texttt{int}) and, crucially, polynomial substitutions.

\subsection{The Three Squares Theorem} \label{sec:three-squares}
The three squares theorem states that for every $n \in \N$ there exist integers $x, y, z \in \Z$ such that $n = x^2 + y^2 + z^2 + z$. It is a consequence of the classical number theory developed in Nathanson's textbook~\cite[Lemma~1.9]{nathanson}. Danilkin and Chevalier~\cite{three-squares-afp} formalized the three squares theorem as well as its number theory prerequisites in our workgroup. Using this result will constitute the last step in our proof. It serves to ultimately replace the single remaining non-negative integer $n \geq 0$ in Lemma~\ref{lem:relation-combining} below by three true integer unknowns.

\subsection{Relation Combining}
The idea of relation-combining is the following: imagine we want to check the two conditions $x > 0, y \ne 0$ with Diophantine polynomials in integer variables. A naive method would use the three-squares formula for each condition, i.e.\ the conditions are true if
\begin{align*}
x &= a^2+b^2+c^2+c+1 \\
y^2 &= d^2+e^2+f^2+f+1
\end{align*}
for some $a,b,c,d,e,f \in \Z$.

However, there is a more efficient method because the two conditions together are equivalent to the single condition
\[
xy^2 = a^2+b^2+c^2+c+1
\;,
\]
using only $a, b, c \in \Z$, cutting the number of integer variables required in half.

This technique has been mastered by Matiyasevich and Robinson with the following result. It combines a divisibility condition, an inequality, and checking whether integers are perfect squares into the existence of only one (!) non-negative integer.

\begin{lemma}[Matiyasevich-Robinson~\cite{MR75}]\label{lem:relation-combining}
For every $q>0$ there is a polynomial $M_q$ with the property that for all integers $A_1,\dots,A_q,R,S,T$ such that $S\neq 0$, the following are equivalent:
\begin{enumerate}
\item $S|T$, $R>0$ and $A_1,\dots,A_q$ are squares,
\item $\exists n \ge 0 : \, M_q(A_1,\dots,A_q,S,T,R,n)=0$.
\end{enumerate}
Specifically, $M_q$ can be given as	
\begin{equation} \label{eq:matiyasevich-polynomial} \begin{split}
&M_q(A_1,\dots,A_q,S,T,R,n)\\
&=\prod_{\varepsilon_1,\dots,\varepsilon_q\in\{\pm1\}^q}\left(S^2n + T^2 - S^2(2R-1)\left(T^2 + X_q^q +\sum_{j=1}^q \varepsilon_j \sqrt{A_j}X_q^{j-1}\right) \right) \\
\end{split} \end{equation}
where $X_q:=1+\sum_{j=1}^{q}A_j^2$.
\end{lemma}

The shorthand notation in equation~\eqref{eq:matiyasevich-polynomial} defines a polynomial because the square roots cancel out in the product (we omit a proof here). In our case, we do not require the general form of this result and instead only its specialization to $q = 3$. Using the automation built into \texttt{poly\_extract} as discussed above, it moreover suffices to simply spell out the explicit function $\texttt{M3} : \Z^7 \rightarrow \Z$. Lemma~\ref{lem:relation-combining} is then formalized as follows.

\begin{lstlisting}[caption=\formalref{Nine\_Unknowns\_N\_Z/Matiyasevich\_Polynomial.thy}{205}.]
lemma relation_combining:
  assumes "S $\neq$ 0"
  shows "(S dvd T $\wedge$ R > 0 $\wedge$ is_square A${}_1$ $\wedge$ is_square A${}_2$ $\wedge$ is_square A${}_3$)
         = ($\exists$n$\geq$0. M3 A${}_1$ A${}_2$ A${}_3$ S T R n = 0)"
\end{lstlisting}

\section{Structure of the Proof}
\begin{figure}
\centering
\includegraphics[page=1]{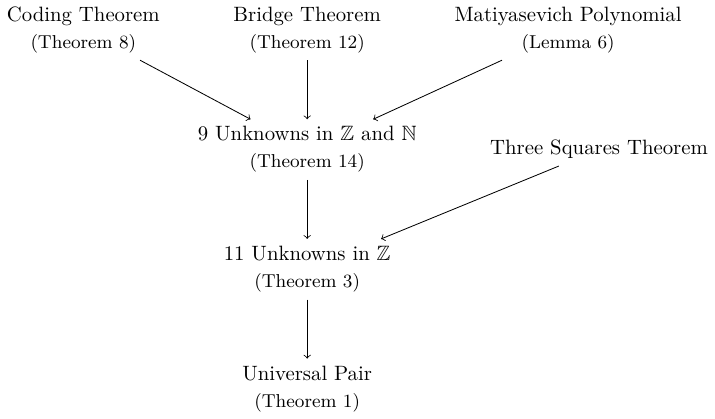}

\caption{Structure of the proof of \cref{thm:11-universal-pair}. Multivariate polynomials are not displayed, they are used throughout.}
\label{fig:proof-structure}
\end{figure}

Our main result \cref{thm:11-universal-pair} is based on three technical statements that capture the core ideas behind universal Diophantine constructions: Coding, Lucas sequences, and relation combining techniques. The dependencies between these results are illustrated in \cref{fig:proof-structure}.

The central challenge in obtaining a universal pair is reducing a general Diophantine equation  --   potentially involving many unknowns  --   to an equivalent equation with a fixed number of unknowns. This reduction is carried out in the three main steps mentioned above. In the process, the original equation is first transformed into intermediate expressions, which are then shown to be equivalent to a fixed Diophantine polynomial.

Crucially, all operations appearing in the intermediate expressions must be Diophantine, that is, expressible using polynomials. Many fundamental mathematical relations admit such representations~\cite{matiyasevich-book}, including divisibility, congruence modulo an integer, and inequalities. Relations involving exponentiation, binomial coefficients and factorials also admit Diophantine representations. However, when deriving universal pairs, it is not enough to establish mere existence; the representation must also be efficient, meaning that it should involve a minimal number of unknowns and remain of low degree.

\section{Coding}
Coding is the key idea that translates a given Diophantine polynomial $P \in \Z[z_1, \ldots, z_\nu]$ with a large number of unknowns into a Diophantine relation of bounded complexity, i.e.\ only requiring a fixed number of unknowns.

\begin{definition}[Code] \label{def:code}
Fix a base $\B = 2^k$ for some integer $k > 1$. Let $z_1, \ldots, z_\nu$ be given and $z_i < \B$ for all $i=1, \ldots, \nu$. We say that $c$ is a \emph{code} for the tuple $(z_1, \ldots, z_\nu)$ with respect to a tuple $(n_1, \ldots, n_\nu)$ of strictly increasing numbers if $c = \sum_{i=1}^\nu z_i \B^{n_i}$.
\end{definition}

If the $z_i$ are the given unknowns of the polynomial $P$ then $c$ consolidates them in a single large number. It is possible to define a number $\alpha_P$ such that the product $K = (1 + c)^{\deg P} \cdot \alpha_P$ contains the value $P(z_1, \ldots, z_\nu)$ in a specific location of its base $\B$ expansion.\footnote{More precisely, such $\alpha_P$ exists if $P$ only contains nonnegative coefficients; the statement can be generalized to arbitrary $P$ by adding an additional correcting term (cf. \cite[Lemma 1.8]{manuscript} where $\alpha_P$ appears as $\coeffs(\B)$).} One can therefore translate the statement $P(z_1, \ldots, z_\nu) = 0$ into a statement about the base $\B$ expansion of $K$. Ensuring that a certain base $\B$ digit of $K$ is zero can then be encoded by counting carries during binary addition. In turn, this condition can be translated into a divisibility condition on a binomial coefficient~\cite[Lemmas~1.3 and~1.4]{manuscript}. 

To formalize codes and work with natural numbers in base $\B$ we make use of our own AFP entry on digit expansions~\cite{afp-digit-expansions}. We also provide a formal definition of the factorial for tuples of natural numbers and formalize the multinomial theorem  --   these are required for working with $\alpha_P$.

\paragraph*{The Coding Theorem}
To combine the various translation steps into a single theorem a comprehensive construction is required. One iteratively defines the auxiliary variables $\mathcal{L}, r, \beta, \alpha, \gamma$ and constructs polynomials $\mathfrak{b}, \mathscr{B}, M, N_0, N_1, N, \mathscr{T}, \Y \in \Z[a, f]$ as well as $c, \mathscr{K}, \mathscr{S}, \mathscr{R}, \X \in \Z[a, f, g]$. All quantities depend on a given polynomial $P \in \Z[a, z_1, \ldots, z_\nu]$ with parameter $a$ and $\nu$ unknowns. In particular, $\K$ corresponds to $K$ given above. We skip the precise definitions here for the sake of brevity (cf. \cite[Definition 2.1]{manuscript}). 

The following statement uses the notation $[j, k) := \{ n \in \N : j \leq n < k \}$.

\begin{theorem}[Coding Theorem~{\cite[Theorem 1]{manuscript}}] \label{thm:coding}
For $a, f \in \N$ and $f > 0$ such that $\bLowercase := \bLowercase(a, f)$ is a power of two, the following are equivalent:
\begin{equation} \label{eq:polynomial-zero}
\exists (z_1,\ldots,z_\nu)\in [0, \bLowercase)^\nu \colon 
P(a, z_1, \ldots, z_\nu) = 0 
\end{equation}
and
\begin{equation} \label{eq:divisibility-binomial}
\exists g \in [\bLowercase, \gamma \bLowercase^\alpha) \colon  \Y(a, f) \, |\, \binom{2\, \X(a, f, g)}{\X(a, f, g)} \;.
\end{equation}
Moreover, \eqref{eq:divisibility-binomial} $\implies $ \eqref{eq:polynomial-zero} already holds when $g \in [0, 2\gamma \bLowercase^\alpha)$.
\end{theorem}

Recall that divisibility, exponentiation and binomial coefficients are Diophantine. We have therefore reduced the original polynomial with $\nu$ unknowns to a Diophantine relation in the parameter $a$ and only two unknowns $f, g$.

Our formalization of this theorem employs two locales. The first locale \texttt{coding\_variables} fixes the polynomial $P$ along with two integers $a$ and $f$, and defines $\delta$ to be the total degree of $P$. Within this locale, the polynomials $\bLowercase, \X,\Y$, as well as the auxiliary variables mentioned above, are introduced. The second locale, \texttt{coding\_theorem}, incorporates the assumptions of \cref{thm:coding}, which are also used in several lemmas leading up to its proof. This structured approach proves advantageous in later stages of the formalization, ensuring that all coding-related concepts are accessed through a unified interface. 
\begin{lstlisting}[caption=\formalref{Coding\_Theorem/Coding\_Theorem.thy}{30}.]
locale coding_theorem = coding_variables + 
  assumes a_nonneg: "a $\geq$ 0"
      and f_pos: "f > 0"
      and b_power2: "is_power2 $\bLowercase$"
      and $\delta$_pos: "$\delta$ > 0"
\end{lstlisting}

Within this locale we formalize two separate statements corresponding to the two directions of the equivalence in \cref{thm:coding}. Here we present the statement of the reverse direction. Because $a$ and $f$ are fixed in the locale, $\X$ now becomes a univariate polynomial in $g$.
\begin{lstlisting}[caption=\formalref{Coding\_Theorem/Coding\_Theorem.thy}{946}.]
theorem coding_theorem_reverse':
  assumes "$\exists$g. 0 $\leq$ g 
            $\land$ g < 2 * $\gamma$ * $\bLowercase$^$\alpha$ 
            $\land$ $\Y$ dvd (2 * $\X$ g choose $\X$ g)"
    shows "$\exists$z. z 0 = a 
            $\land$ ($\forall$i. 0 $\leq$ z i $\land$ z i < $\bLowercase$) 
            $\land$ insertion z P = 0"
\end{lstlisting}

A significant portion of the proof of this statement involves establishing bounds between the various variables and polynomials. The inequalities originally stated in our mathematical draft were often not accurate or missing crucial assumptions. It was convenient to then work out the details in Isabelle which tracked precisely which bounds had already been established and under which conditions  --   for this type of mathematics, the proof \emph{assistant} lived up to its name! We made heavy use of literal facts (writing \texttt{\guilsinglleft b > 0\guilsinglright} instead of using an identifier like \texttt{b\_gt\_0}) which additionally improved the presentation of the proof (cf. the file \texttt{Lower\_Bounds.thy}).

\section{Lucas Sequences and the Pell Equation}
The Coding Theorem introduces an exponential relation  --   namely that the base $\bLowercase$ be a power of two  --   and a binomial coefficient. Both need to be converted into an equivalent polynomial condition. To do so efficiently, we rely on the solutions of a Pell equation, also known as Lucas sequences. 

\begin{definition}[Lucas sequence]
A \emph{Lucas sequence with parameter $A\in\Z$} is a sequence $(x_n)_{n\in \Z}$ that satisfies the recursive relation 
\begin{equation}\label{eq:lucas-general-recurrence}
x_{n+1}=Ax_n-x_{n-1}.
\end{equation}
Each such sequence is defined uniquely by two initial values, and we define two particular Lucas sequences with initial values $\psi_0(A)=0, \psi_1(A)=1$ as well as $\chi_0(A)=2, \chi_1(A)=A$.
\end{definition}
In Isabelle, we first define these Lucas sequences for $n \in \N$.
\begin{lstlisting}[caption=\formalref{Lucas\_Sequences/Lucas\_Sequences.thy}{7}.]
fun $\psi$ :: "int $\Rightarrow$ nat $\Rightarrow$ int" where
  "$\psi$ A 0 = 0" | "$\psi$ A (Suc 0) = 1" |
  "$\psi$ A (Suc (Suc n)) = A * ($\psi$ A (Suc n)) - ($\psi$ A n)"

fun $\chi$ :: "int $\Rightarrow$ nat $\Rightarrow$ int" where
  "$\chi$ A 0 = 2" | "$\chi$ A (Suc 0) = A" |
  "$\chi$ A (Suc (Suc n)) = A * ($\chi$ A (Suc n)) - ($\chi$ A n)"
\end{lstlisting}
We then extend these recursive definitions to all $n \in \Z$ by using the symmetries of these sequences, given by $\psi_{-n}(A) = -\psi_n(A)$ and $\chi_{-n}(A) = \chi_n(A)$. With these properties, it follows directly that Equation~\eqref{eq:lucas-general-recurrence} holds for the entire integer-indexed sequence.

\begin{lstlisting}[caption=\formalref{Lucas\_Sequences/DFI\_square\_1.thy}{386}.]
definition $\psi$_int :: "int $\Rightarrow$ int $\Rightarrow$ int"
  where "$\psi$_int A n = (if n $\geq$ 0 then 1 else -1) * $\psi$ A (nat (abs n))"

definition $\chi$_int :: "int $\Rightarrow$ int $\Rightarrow$ int"
  where "$\chi$_int A n = $\chi$ A (nat (abs n))"
\end{lstlisting}

These two second-order recurrences grow exponentially: for example, if $n \ge 2$ and $A>1$, we have $(A-1)^n < \psi_{n+1}(A) < A^n$. This is the property we will use to encode that $\bLowercase$ is a power of two. Importantly, these recurrences are Diophantine, in the sense that they are solutions of the following polynomial equation.

\begin{lemma}[Pell's equation]
\label{lem:PellEquation}
If $d=A^2-4$, the Pell equation
\begin{equation}
X^2-dY^2=4
\end{equation}
has the property that its solutions are exactly the pairs $(X,Y)=(\chi_n(A),\psi_n(A))$ for $n\in\Z$.
\end{lemma}
The following formal statements are sufficient for our purposes. 

\begin{lstlisting}[caption=\formalref{Lucas\_Sequences/DFI\_square\_1.thy}{582}.]
lemma lucas_solves_pell: "(A${}^2$-4)*($\psi$_int A m)${}^2$ + 4 = ($\chi$_int A m)${}^2$"
lemma pell_yields_lucas: "($\exists$k. (A${}^2$-4)*Y${}^2$ + 4 = k${}^2$) = ($\exists$m. Y = $\psi$_int A m)"
\end{lstlisting}
In the course of the proofs of these statements, totaling about 2,000 lines, we formalize many auxiliary properties of Lucas sequences. This includes both a wealth of elementary facts (cf.~\cite[Lemma~3.4]{manuscript}) and less trivial properties (cf.~\cite[Lemmas~3.7--3.12]{manuscript}). All of the latter had previously been stated or sketched in the literature, but this formalization, and our associated manuscript, contain their complete proofs for the first time.

\paragraph*{A Jungle of Variables to Build a Bridge}

The specific relations which will represent the exponential conditions from the previous section (power of two and binomial coefficient) rely on a dozen auxiliary definitions. This is not the only list of this length; also the Coding Theorem~\ref{thm:coding} as well as the final construction in Theorem~\ref{thm:9-unknowns-ZN} depend on comparable jungles of variables. We give Definition~\ref{def:VariableDefinition} here as an example that illustrates this complexity.

It is only partially possible to assign high-level meaning to the variable definitions below. Note that if $D$ is a square, then \cref{eq:Def_D} corresponds to a Pell equation with $d = A^2 - 4$. Moreover, the lemma \texttt{pell\_yields\_lucas} can be applied to relation~\eqref{eq:UVK} to deduce that $J$ is the value of a Lucas sequence with parameter $U^2V$. Many of the remaining variables serve primarily as scaffolding, ensuring the necessary conditions to apply the appropriate number-theoretic lemmas.

\begin{definition}
\label{def:VariableDefinition}
Given variables $X,Y,b,g,h,k,l,w,x,y$ we define the following variables:
\begin{subequations}\label{eqs:VariableDefinition}
\begin{align}
U &:= 2lXY \label{eq:Def_U} \\
V &:= 4gwY \label{eq:Def_V} \\
A &:= U(V+1) \label{eq:Def_A} \\
B &:= 2X+1 \label{eq:Def_B} \\
C &:= B+(A-2)h \label{eq:Def_C} \\
D &:= (A^2-4)C^2+4 \label{eq:Def_D} \\
E &:= C^2Dx \label{eq:Def_E} \\
F &:= 4(A^2-4)E^2+1 \label{eq:Def_F} \\
G &:= 1+CDF-2(A+2)(A-2)^2E^2 \label{eq:Def_G} \\
H &:= C+BF+(2y-1)CF \label{eq:Def_H} \\
I &:= (G^2-1)H^2+1 \label{eq:Def_I} \\
J &:= X+1+k(U^2V-2) \label{eq:Def_J} \\
S &:= 2A - 5 \label{eq:Def_S} \\
T &:= 3bwC-2(b^2w^2-1) \label{eq:Def_T}
\end{align}
\end{subequations}
Denote the set of squares by $\square$. We define the following relations on the just-constructed variables:
\begin{subequations}\label{eqs:thm2-properties}
\begin{align}
&DFI \in \square \label{eq:DFI} \\
&(U^{4}V^2-4)J^2+4\in\square \label{eq:UVK} \\
&S \,\big|\, T \label{eq:pAppWW_new}\\
& \left(\frac{C}{J}-lY\right)^2 < \frac{1}{16g^2} \label{eq:g}
\;.
\end{align}
\end{subequations}
\end{definition}
Notice that this list contains exactly those relations which can be encoded by the Matiyasevich polynomial in Lemma~\ref{lem:relation-combining}, namely multiple perfect squares, one divisibility condition, and one inequality. This is what earns our most technical theorem its name, for it bridges the gap between the Coding Theorem~\ref{thm:coding} and the Matiyasevich polynomial.

\begin{theorem}[Bridge Theorem~{\cite[Theorem 2]{manuscript}}] \label{thm:lucas}
Given integers $b\ge 0$, $X\geq 3b$, $Y\geq \max \{ b,2^8 \}$ and $g\ge 1$, the following are equivalent:
\begin{enumerate}
\item \label{proofitem:Binom} $b$ is a power of 2 and $Y | \binom{2 X}{X}$;
\item \label{proofitem:Dioph2} $\exists h,k,l,w,x,y\geq b$ such that the Diophantine conditions \eqref{eq:DFI}, \eqref{eq:UVK}, \eqref{eq:pAppWW_new}, \eqref{eq:g} hold.
\end{enumerate}
Moreover, the implication \eqref{proofitem:Binom} $\implies$ \eqref{proofitem:Dioph2} holds already if $X\ge b$ (rather than $X\ge 3b$).
\end{theorem}

\begin{remark}
The slightly simplified version of Theorem~\ref{thm:lucas} stated here illustrates the role of this theorem for the big picture. In full detail (cf.~\cite[Theorem~2]{manuscript}), it establishes the equivalence of three statements, where the third statement is a slightly modified version of \cref{proofitem:Dioph2}. 
\end{remark}

This statement is formalized in Isabelle as presented below. The lists of equations~\eqref{eqs:VariableDefinition} and \eqref{eqs:thm2-properties} are implemented in the locale \texttt{bridge\_variables}. Notice that all relations are implicitly functions of the variables $X,Y,b,g,h,k,l,w,x,y$ listed in Definition~\ref{def:VariableDefinition}. To ease legibility below, we color these variables in gray whenever they are arguments to another function.

\begin{lstlisting}[caption=\formalref{Bridge\_Theorem/Bridge\_Theorem.thy}{14}.,escapechar={|}]
definition (in bridge_variables) bridge_relations where
  "bridge_relations k l w h x y b g Y X $\equiv$
      is_square (D |\color{gray}l w h g Y X| * F |\color{gray}l w h x g Y X| * I |\color{gray}l w h x y g Y X|)
      $\wedge$ is_square ((U |\color{gray}l X Y|^4 * V |\color{gray}w g Y|^2 - 4) * J |\color{gray}k l w g Y X|^2 + 4)
      $\wedge$ S |\color{gray}l w g Y X| dvd T |\color{gray}l w h g Y X b|
      $\wedge$ (4*g*C |\color{gray}l w h g Y X| - 4*g*l*Y*J |\color{gray}k l w g Y X|)^2 < (J |\color{gray}k l w g Y X|)^2"

theorem (in bridge_variables) bridge_theorem_simplified:
  fixes b Y X g :: int
  assumes "b $\geq$ 0" and "Y $\geq$ b" and "Y $\geq$ 2^8" and "X $\geq$ 3*b" and "g $\geq$ 1"
  shows "(is_power2 b $\wedge$ Y dvd (2 * X choose X))
            = ($\exists$h k l w x y :: int. bridge_relations k l w h x y b g Y X
                $\wedge$ (h$\geq$b) $\wedge$ (k$\geq$b) $\wedge$ (l$\geq$b) $\wedge$ (w$\geq$b) $\wedge$ (x$\geq$b) $\wedge$ (y$\geq$b))"
\end{lstlisting}

\section{Eleven Unknowns in \texorpdfstring{\boldmath $\Z$}{Z}}

In the previous sections, we have detailed multiple translation theorems. \Cref{thm:coding} translates between an arbitrary polynomial $P$ having a (parametric) solution and a Diophantine predicate involving exponentiation and a binomial coefficient. \Cref{thm:lucas} translates exponentiation and the binomial coefficient into more elementary Diophantine relations. Finally, \Cref{lem:relation-combining} allows us to efficiently represent precisely these relations through a single polynomial. 

\paragraph*{An Intermediate Theorem}
Before we can prove our two main Theorems~\ref{thm:11-universal-pair} and~\ref{thm:11-unknowns} stated in the introduction, we first combine the previous statements into an intermediate result. Let a polynomial $P$ with natural number unknowns be given. Our goal is to construct a polynomial $Q$ with a fixed number of unknowns which should have a solution whenever $P$ has a solution. 

We do not give the elaborate definition of $Q$ here; it specifies how the various polynomials defined previously are substituted into each other~\cite[Definition 5.2]{manuscript}. In our construction, $Q$ has nine unknowns, eight of which are integers and the remaining one is a natural number. One can then prove the following theorem:

\begin{theorem}[9 Unknowns in $\Z$ and $\N$~{\cite[Theorem 3]{manuscript}}] \label{thm:9-unknowns-ZN}
For all $a \in \N$ the following are equivalent: 
\begin{gather} \label{eq:P-solution}
\exists z_1, \ldots, z_\nu \in \N \colon 
P(a, z_1, \ldots, z_\nu) = 0 \\
\exists f, g, h, k, l, w, x, y \in \Z, n \in \N \colon 
Q(a, f, g, h, k, l, w, x, y, n) = 0 \;. 
\label{eq:Q-solution}
\end{gather}
\end{theorem}

\enlargethispage{1.1\baselineskip}
The proof of this theorem proceeds by combining \cref{thm:coding,thm:lucas,lem:relation-combining}. We structure the formalization of \cref{thm:9-unknowns-ZN} into two Isabelle theories: the first one contains all definitions leading up to $Q$ and the second one implements the proof of the theorem itself. When defining $Q$ we make frequent use of the \texttt{poly\_extract} command.

We formalize two versions of the statement of \cref{thm:9-unknowns-ZN}. In the first version, $Q$ is simply an Isabelle term. The second version presented below uses $Q$ of type \texttt{int mpoly}. Note the use of \texttt{is\_nonnegative} (defined in the obvious way) which we use because $P$ has natural number unknowns. The syntax \lstinline{z(0 := a)} is used to substitute the parameter $a$ into the first variable of \texttt{P}. 
\begin{lstlisting}[caption=\formalref{Nine\_Unknowns\_N\_Z/Nine\_Unknowns\_N\_Z.thy}{507}.]
theorem nine_unknowns_over_Z_and_N: 
  fixes P :: "int mpoly"
  shows "($\exists$z :: nat $\Rightarrow$ int. is_nonnegative z $\land$ 
              insertion (z(0 := int a)) P = 0) 
       = ($\exists$f g h k l w x y n. n $\geq$ 0 $\land$
              insertion ((!) [int a, f, g, h, k, l, w, x, y, n])
              (combined_variables.Q_poly P) = 0)"
\end{lstlisting}

\paragraph*{Transforming the Last Natural Number Unknown}
The statement of \cref{thm:9-unknowns-ZN} is close to that of \cref{thm:11-unknowns}, the remaining step being that $Q$ still has one natural number unknown which needs to be converted into integer unknowns. One can introduce additional unknowns $n_1, n_2, n_3 \in \Z$ and replace $n$ with $n_1^2 + n_2^2 + n_3^2 + n_3$ to obtain a polynomial $\tilde Q$:
\[ \tilde Q (a, f, g, h, k, l, w, x, y, n_1, n_2, n_3) = Q(a, f, g, h, k, l, w, x, y, n_1^2 + n_2^2 + n_3^2 + n_3) \]
We can use this to prove \cref{thm:11-unknowns}.
\begin{proof}[Proof of \cref{thm:11-unknowns}]
By the three squares Theorem, any $n \geq 0$ can be represented as $n_1^2 + n_2^2 + n_3^2 + n_3$. Therefore, $\tilde Q$ has a solution whenever $n \geq 0$ and $Q$ has a solution, i.e.\ when \eqref{eq:Q-solution} holds. Using \cref{thm:9-unknowns-ZN}, we can thus deduce that $P$ has a solution whenever $\tilde Q$ has a solution. The preceding construction uses exactly 11 unknowns, completing the proof of~\cref{thm:11-unknowns}.
\end{proof}

To formally prove \cref{thm:11-unknowns}, our formalization implements the combination of \cref{thm:9-unknowns-ZN} and the three squares theorem in an additional 50 lines of Isabelle code. The formal statement assumes \lstinline{is_diophantine_over_N_with A P}, that is, $A \subset \N$ is a Diophantine set and represented by the polynomial $P$. Then, by definition, $a \in A$ is equivalent to $\exists z_1, \ldots, z_\nu \in \Z^\nu. \; P(a, z_1, \ldots, z_\nu) = 0$.

\begin{lstlisting}[breaklines=true,caption=\formalref{Eleven\_Unknowns\_Z.thy}{168}.]
lemma eleven_unknowns_over_Z_polynomial: 
  fixes A :: "nat set" and P :: "int mpoly"
  assumes "is_diophantine_over_N_with A P"
  shows "a $\in$ A = ($\exists$z1 z2 z3 z4 z5 z6 z7 z8 z9 z10 z11.
                  insertion ((!) [int a, z1, z2, z3, z4, z5, z6, z7, z8, z9, z10, z11]) (Q_tilde_poly P) = 0)"
\end{lstlisting}

\bigskip

Using the polynomial $\tilde Q$ it is also possible to derive a universal pair, i.e.\ prove our \cref{thm:11-universal-pair}.

\begin{proof}[Proof of \cref{thm:11-universal-pair}]
Let $(\nu, \delta)_\N$ be a universal pair. Therefore, every polynomial $P$ can be represented in at most this complexity. Theorem~\ref{thm:11-unknowns} establishes that $\tilde Q$ always has eleven integer unknowns, independent of $P$. Calculating the degree of $\tilde Q$ now gives the function $\eta(\nu, \delta)$ as claimed~\cite[Section~6]{manuscript}. Assuming the universal pairs $(58, 4)_\N$ and $(32, 12)_\N$, one can calculate the explicit value of $\eta$ and obtains that 
\[
(11, 1.68105\cdot 10^{63}) \qquad \text{and} \qquad (11, 9.50818 \cdot 10^{53}) 
\]
are universal. 
\end{proof}

To formally prove \cref{thm:11-universal-pair}, we invoke our tool \texttt{poly\_degree Q\_tilde\_poly} and combine the result of the automated computation with some manual evaluation and simplification. This ultimately computes the function $\eta(\nu, \delta)$ which we define in Isabelle exactly as defined in \cref{thm:11-universal-pair}.

The definition of universal pairs over the natural numbers is formalized as follows. Mirroring our mathematical notation, we introduce the custom syntax \texttt{($\nu$, $\delta$)${}_\mathbb{N}$} in Isabelle.
\begin{lstlisting}[caption=\formalref{Universal\_Pairs.thy}{5}.]
definition universal_pair_N ("(_, _)${}_\mathbb{N}$") where
  "universal_pair_N $\nu$ $\delta$ $\equiv$ ($\forall$A::nat set. is_diophantine_over_N A $\rightarrow$
        ($\exists$P::int mpoly. max_vars P $\leq$ $\nu$ $\wedge$ total_degree P $\leq$ $\delta$ $\wedge$
                        is_diophantine_over_N_with A P))"
\end{lstlisting}

An analogous definition is made for \texttt{universal\_pair\_Z}, which introduces the custom syntax \texttt{($\nu$, $\delta$)${}_\mathbb{Z}$}.

At last, \cref{thm:11-universal-pair} can be stated concisely.
\begin{lstlisting}[caption=\formalref{Universal\_Pairs.thy}{15}.]
theorem universal_pairs_Z_from_N: "($\nu$, $\delta$)${}_\mathbb{N}$ $\Longrightarrow$ (11, $\eta$ $\nu$ $\delta$)${}_\mathbb{Z}$"
\end{lstlisting}

The specific universal pairs stated above are now just special cases of this theorem. For example, the second of the two pairs (the smaller one) reads:
\begin{lstlisting}[caption=\formalref{Universal\_Pairs.thy}{76}.]
theorem universal_pair_2:
  assumes "(32, 12)${}_\mathbb{N}$" 
  shows "(11, 950817549694171759711025515571236610412597656252821888)${}_\mathbb{Z}$"
\end{lstlisting}

\section{Conclusion}

We have formalized a universal Diophantine construction that implies a stronger version of Hilbert's Tenth Problem. Already the pre-print~\cite{manuscript} of our novel Theorem~\ref{thm:11-universal-pair} thus appears with the certificate of correctness presented herein. This result has been largely developed in parallel to its Isabelle formalization. In the course of the project, we also built a substantial library of advanced facts about multivariate polynomials and formalized results from a textbook on number theory. The total development consists of 20,000 lines of code.

\paragraph*{Experiences from an In-situ Formalization}
The formalization was executed by a student workgroup at a time when only a very early draft of the underlying mathematics was available. Hence we say that this draft manuscript was formalized \emph{in-situ}, in its natural environment --  a mathematics department --  , and before a version of publication quality had been prepared.

Interacting with a proof assistant significantly shaped the development of the mathematics and the mathematical text: we refined arguments, removed or weakened redundant assumptions, and corrected a large number of minor bugs throughout our original draft. Even our theorems benefited significantly from the formalization. In fact, it was not before we completed its formalization that we managed to obtain the correct formula for $\eta(\nu, \delta)$, now given in \cref{thm:11-universal-pair}. Also the first pre-print of this article still contained a wrong formula.

To give another example, we had also overlooked that one of our technical lemmas depended on a lemma in the literature~\cite[Lemma 10]{SunDFI}. Without noticing our initial fallacious proof of the technical lemma, the team responsible for its formalization had rediscovered a correct proof on their own. This subtle but severe omission in our original proof was discovered only during the final review phase of our manuscript. We then used our already complete formalization to write a correct pen-and-paper proof. Thanks to the structured proof language of Isabelle, even our collaborators who had no experience with Isabelle were able to easily read the formal proof.

Besides those two major mistakes, we found that for many technical parts of the proof, Isabelle functioned as a true proof \textit{assistant}: supporting us authors by keeping track of assumptions and bounds so that we could focus on the more conceptual elements of the proof. For a complete account of the benefits and challenges of this unusual method of doing mathematics, we refer to the extended reflection in our full article~\cite[Section~7]{manuscript}.

\paragraph*{Outlook}
Future research could extend this work by formalizing universal pairs over $\N$ as constructed in the article by Jones~\cite{9var}. We believe that the infrastructure we have built in this project would provide a strong starting point for such work, eliminating the need to formalize further prerequisites on multivariate polynomials. 

Since this project bridges the frontier between mathematical research and formalization in a proof assistant, further progress will naturally also involve new mathematical discoveries. Hilbert's Tenth Problem has been proven to be solvable for $\delta \leq 2$ or alternatively $\nu = 1$~\cite{9var}, but remains largely open for $\delta = 3$ and for the whole range $2 \leq \nu \leq 10$.

\enlargethispage{-\baselineskip}
\bibliography{references}

\end{document}